\documentclass{article}
\usepackage{spconf,amsmath,graphicx}
\usepackage{amsmath}
\usepackage{amssymb}
\usepackage{subcaption}
\usepackage{graphicx} 
\usepackage{multirow}
\usepackage{array}
\usepackage{booktabs}
\usepackage{courier}
\usepackage{float}

\title{Similarity measures for vocal-based drum sample retrieval using deep convolutional auto-encoders}

\name{Adib Mehrabi, Keunwoo Choi, Simon Dixon, Mark Sandler}
\address{Queen Mary University of London, London, UK\\
Centre for Digital Music, EECS\\
E1 4FZ, London, UK\\
\texttt{\{a.mehrabi, keunwoo.choi\}{@}qmul.ac.uk}}

\begin{document}
\ninept

\maketitle

\begin{abstract}
The expressive nature of the voice provides a powerful medium for communicating sonic ideas, motivating recent research on methods for query by vocalisation. Meanwhile, deep learning methods have demonstrated state-of-the-art results for matching vocal imitations to imitated sounds, yet little is known about how well learned features represent the perceptual similarity between vocalisations and queried sounds. 
In this paper, we address this question using similarity ratings between vocal imitations and imitated drum sounds. We use a linear mixed effect regression model to show how features learned by convolutional auto-encoders (CAEs) perform as predictors for perceptual similarity between sounds. Our experiments show that CAEs outperform three baseline feature sets (spectrogram-based representations, MFCCs, and temporal features) at predicting the subjective similarity ratings. We also investigate how the size and shape of the encoded layer effects the predictive power of the learned features. The results show that preservation of temporal information is more important than spectral resolution for this application.

\end{abstract}


\begin{figure*}[htp]
\begin{center}
\includegraphics[width=2\columnwidth]{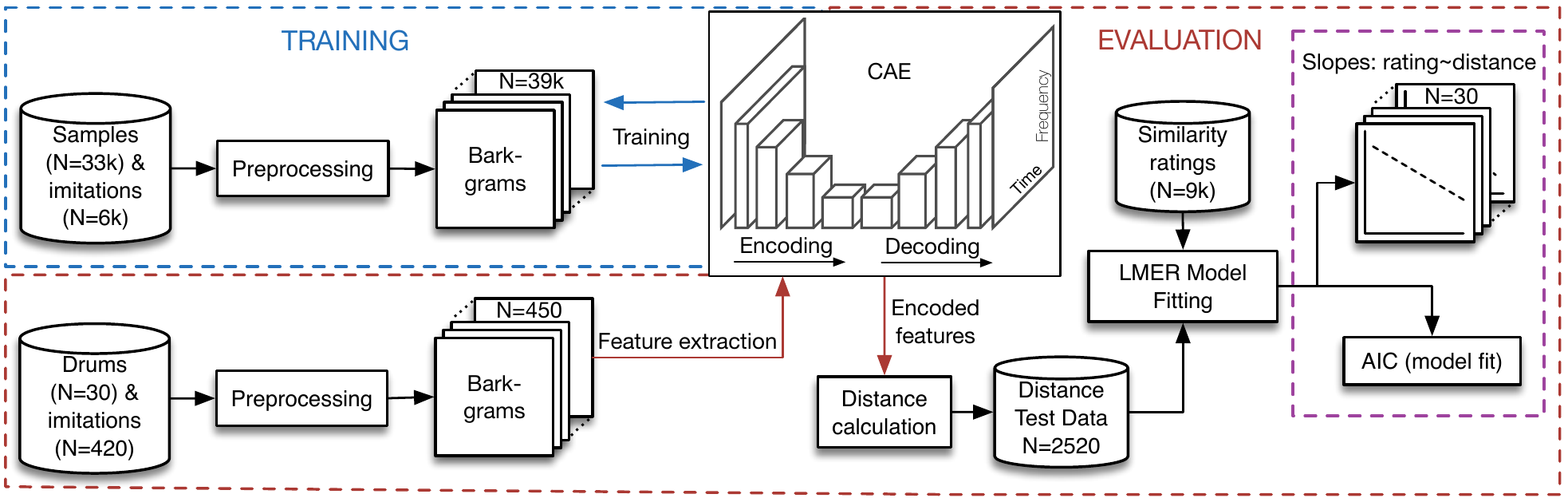}
\caption{Overview of the complete work flow. All audio (training and test data) is preprocessed to create 128x128 barkgram representations. The trained CAE is used to extract features from the test data. Euclidean \textit{distance} between each imitation and its imitated sound is then computed, and fitted with the \textit{rating} data to an LMER model. Performance of the 14 feature sets (3 baselines and 11 CAE networks) is measured by 1) AIC for model fit, and 2) the proportion of imitated sounds that have a significantly negative slopes for $rating \sim\ distance$.}
\label{fig:block}
\end{center}
\end{figure*}

\section{Introduction and Related Work}
\label{sec:intro}

Searching for audio samples is a core part of the electronic music making process, yet is a time consuming task, and a key area for future technological development \cite{andersen2015giantsteps}. This task typically involves browsing lists of badly labelled files, relying on filenames such as `big\_kick' or `hi-hat22'. Such methods for browsing sound libraries limit the users' ability to efficiently find the sounds they are looking for. Meanwhile, the voice provides an attractive medium for effectively communicating sonic ideas \cite{lemaitre2011vocal, lemaitre2014effectiveness}, as it can be used to express timbral, tonal and dynamic temporal variations \cite{sundberg1990science}. Moreover, previous research demonstrates that musicians are able to accurately vocalise important acoustic features of musical sounds \cite{lemaitre2016vocal, mehrabi2017vocal}. 

Query by vocalisation (QBV) is the process of searching for sounds based on vocalised examples of the desired sound. Typically, QBV systems extract audio features from a vocalisation, which can then be  compared to the features of sounds in a sample library (to return class labels or a ranked list of sounds). Initial approaches to QBV used heuristic based features \cite{blancas2014sound, roma2015querying}. Morphological features describing the high-level temporal evolution of sounds have also been applied to QBV \cite{marchetto2015morphological}, however drum sounds generally have similar high-level temporal morphology (i.e.\ rise-fall), so these types of features are less applicable here. 

Recent work has shown that features learned using stacked auto-encoders (SAEs) outperform heuristic descriptors such as MFCCs (Mel-frequency cepstral coefficients) for QBV tasks. SAEs utilise a deep learning structure where multiple layers learn an efficient representation to encode the input. These have been applied in 2 QBV scenarios: supervised learning, using the features to train a classifier \cite{zhang2015retrieving}; and unsupervised search, based on distance between sounds in a Euclidean feature space \cite{zhang2016, zhang2016supervised}. Furthermore, in \cite{zhangiminet} the authors present a QBV system based on convolutional neural networks (CNNs) implemented in a semi-Siamese network structure. In this case the convolutional layers are trained to learn feature representations from constant-Q spectrograms of vocal imitations and the imitated sounds. The CNN is followed by fully connected layers to match input vocalisations to audio samples, requiring each sample in a sound library to be compared to a vocal query. The system shows promising results for matching vocal imitations to the imitated sounds, however in the general case QBV systems require efficient, deployable querying. Using this method, a single query on a dataset with $N$ data samples requires $N$ forward-pass computations of the network, which is significantly demanding, for example compared to nearest neighbour search in a feature vector space.

Whilst both SAE and CNN approaches show promising performance in terms of retrieving an \textit{imitated} sound from a set of audio samples, none of the above mentioned QBV methods consider the \textit{perceptual similarity} between the query and retrieved sounds. Central to the evaluation of these approaches is the assumption that the target sound is indeed the sound that was imitated, and the task is to match the imitations and imitated sounds accordingly. However, we consider a use case in which the query is not necessarily an imitation of a sound in the database, and investigate which feature representations correlate well with the perceptual similarity between an imitation and a set of audio samples. 

In this paper we evaluate the performance of both heuristic and learned features for QBV of drum sounds. An overview of our approach is illustrated in Fig.~\ref{fig:block}. We present a set of convolutional auto-encoders (CAEs) trained on a dataset of $\sim\ $33k audio samples and $\sim\ $6k vocalisations. These are used to extract features from 420 vocal imitations of 30 drum sounds. The feature sets are evaluated using perceptual similarity ratings between the vocal imitations and the imitated drum sounds. We include 4 types of features: (1) a spectrogram based representation from \cite{pampalk2008}, which the authors show to correlate strongly with perceptual similarity between drum sounds; (2) MFCCs; (3) temporal descriptors; (4) encoded representations from the CAEs. We compare 11 CAEs, which differ in both the size of the encoded feature tensor and the shape of the encoded layer in the temporal and spectral dimensions.  

\section{Problem Definition}

The task is to establish which audio features best correlate with  perceptual similarity between real drum sounds (the \textit{imitated sounds}) and vocal imitations of drum sounds (the \textit{imitations}). Specifically, we are interested in \emph{i)} how heuristic descriptors perform compared to learned features using CAEs, and \emph{ii)} the importance of temporal vs.~spectral dimensions and the size of the encoded tensors from the CAEs. We limit the problem to a set of 30 drum sounds: 6 from each of 5 classes (kick, snare, cymbal, hi-hat, tom-tom), and consider only the similarity between imitations and within-class sounds (e.g.\ between the imitation of a snare and the actual snare sounds). 

\section{Experiments}

\subsection{Baseline Methods}

We use 3 baseline methods. The first (PK08) is a spectrogram-based measure of similarity from \cite{pampalk2008}. This has been shown to correlate highly with perceptual similarity ratings between within-class drum sounds, and we are interested in how well it transfers to our application. In summary, similarity between 2 sounds is measured as the Euclidean distance between their vectorised barkgrams, constructed from a spectrogram with the following parameters: 93ms window; 87.5\% overlap; Bark scale (72 bins); loudness in dB and scaled using Terhardt's ear model \cite{terhardt1979calculating}. The barkgrams are time-aligned, and where 2 sounds are not of the same length the shorter is zero padded to the length of the longer one. 

For the second method (MFCC) we calculate the first 13 MFCCs for each sound (excluding MFCC 0) with first and second order derivatives, using a 93ms time window and 87.5\% overlap. The mean and variance of each MFCC and its derivatives are calculated for each sound, yielding 78 features. The third method (TEMP) is a set of 5 temporal features: log attack time (LAT); temporal centroid (TC); LAT/TC ratio; temporal crest factor (TCF); and duration. We calculate LAT and TC as per the definition in \cite{pampalk2008}. TCF is calculated over the entire time domain signal (rectified), and is the maximum value divided by the root mean squared. 

\subsection{CAE Networks}
\vspace{-3pt}

\subsubsection{Model Architecture} 
The basic architecture is a CAE with four 2D convolution layers in its encoder/decoder. Each convolutional layer is followed by batch normalisation and ReLU activation layers. To avoid checker board artefacts caused by deconvolution layers \cite{odena2016deconvolution} we apply upsampling prior to each decoding convolutional layer. As such, each decoding deconvolution layer is an upsampling layer followed by a 2D convolution layer with $(1, 1)$ stride. We vary the kernel size of the first and last layers while using fixed $(10, 10)$ kernels for the other convolution layers. The encoding layers have [8, 16, 24, 32] kernels (layers 1-4 respectively) which is mirrored in the decoder, i.e., [32, 24, 16, 8]. A single-channel convolution layer is used for the output layer. The activation of the last layer of the encoder is flattened and taken as the feature vector for a given test sample.  

The kernel size and stride of the convolution (or upsampling) layers are varied in order to compare the shape (i.e.\ square, wide, tall) and size of the encoded representation, respectively. Details for 11 variants of the above model are given in Table \ref{tab:results}. 

\subsubsection{Training Data and Pre-processing}
\label{sec:trainingdata}
The network is designed to learn a broad range of vocal and percussion related sounds including  \textit{i)} short, percussive/non-percussive and pitched/unpitched sounds, and \textit{ii)} non-verbal vocalisations. The training dataset is made up of 24,294 percussion sounds, 4,884 sound effects and 4,523 single note instrument samples. In addition, we include 4,429 vocal imitations of instruments, synthesisers and everyday sounds from \cite{cartwright2015vocalsketch}, and 1,387 vocal imitations of 72 short synthesised sounds from \cite{mehrabi2017vocal}. This results in a dataset of $\sim\ $39k sounds, of which $\sim\ $6k are vocal imitations.

For each sound in the training set we compute the barkgrams from spectrograms with a 93 ms time window and 87.5\% overlap, using 128 Bark bins. As with the PK08 baseline, the magnitudes are scaled (in dB) using Terhardt's ear model curves \cite{terhardt1979calculating}. To achieve a fixed size representation for all sounds, we either zero-pad or truncate the barkgrams to 128 frames ($\approx$ 1.5 seconds). 

\subsubsection{Training Procedure}
The models are implemented using Keras \cite{chollet2015keras} and Tensorflow \cite{abadi2016tensorflow}. Training and validation sets are 70:30\% split from the training data (Section \ref{sec:trainingdata}). As the training dataset contains 5.5 times more audio samples than vocal imitations, and we are equally interested in learning both sound types, we specify a 50/50\% split of audio samples/vocal imitations for each batch (128 data samples). The models are all fitted using the Adaptive Moment estimation (Adam) optimiser \cite{kingma2014adam} with a learning rate of 0.001, and mean squared error loss function. We use the early-stopping scheme for no improvement in validation loss after 10 epochs. The best (i.e.\ lowest validation loss) model for each parameter setting is selected for the analysis. 

\section{Evaluation}

\subsection{Test data}

\textbf{The 30 drum sounds} were taken from the fxpansion\footnote{https://www.fxpansion.com} \emph{BFD3 Core} and \emph{8BitKit} sample libraries, which include a range of acoustic and electronic drum samples. Vocal imitations of each sound were recorded by 14 musicians ($>$5 years experience), giving 420 imitations. The recordings took place in an acoustically treated room at the Centre for Digital Music, Queen Mary University of London\footnote{http://www.eecs.qmul.ac.uk/facilities/view/control-room}.

\textbf{Perceptual similarity ratings} between the imitations and each of the within-class drum sounds were collected from 63 listeners via a web based listening test, using a format based on the MUSHRA protocol for subjective assessment of audio quality \cite{recommendation20031534}. Whilst the MUSHRA standard specifies the use of expert listeners, it has recently been shown that lay listeners can provide comparable results to experts for measuring audio quality \cite{cartwright2016fast}. Each listener was presented with 30 tests. For each test the listener was presented with a (randomly selected) vocal imitation and the 6 within-class drum sounds (one being the imitated sound). The listener then rated the similarity between the imitation and each drum sound (giving 6 similarity ratings per test), on a continuous scale from `less similar' to `more similar'. 

Of the 30 test pages, 28 were unique and 2 were random duplicates. These were included for post-screening of the listeners, as recommended in the MUSHRA standard \cite{recommendation20031534}. Listener reliability was assessed using the Spearman rank correlation between the two duplicate test pages for each listener. We considered reliable listeners as those who were able to replicate their responses for at least one of the duplicates with $\rho >= 0.5$, i.e.\ large positive correlation \cite{cohen1977statistical}. There were 51 reliable listeners, for whom $\rho$ = 0.63/0.04 (mean/standard error), giving 9,126 responses from 1521 tests (excluding duplicates). We then computed Kendall's coefficient of concordance, $W$ \cite{kendall1939problem} on the ranked responses for each imitation. The mean/standard error of $W$ = 0.61/0.01, indicating moderate to strong agreement amongst the reliable listeners \cite{schmidt1997managing}.

Analysis of the ratings indicated that listeners were able to correctly identify the imitated sound with above chance accuracy (37\% of cases, chance = 16\%), and the imitated sound was rated first or second most similar to the imitation in 60\% of tests. This indicates that although the imitations were often rated as being most similar to the imitated sounds, there are a considerable number of cases (up to 40\%) where 2 of the 6 within-class sounds were rated more similar to the imitation than the imitated sound. This highlights the potential importance of perceptual similarity measures for tasks such as QBV, depending on whether the task is to identify and return an \textit{imitated} sound, or to return the \textit{most similar} sound. The 9126 similarity ratings are used as as a ground truth from which to measure the performance of each of the feature sets. 

\begin{table*}[htp]
\fontsize{7}{7}\selectfont
\centering
\begin{tabular}{lllllllllllll}
\toprule
\multirow{2}{*}{Type} & \multirow{2}{*}{Feat. set} & \multirow{2}{*}{L1/8 kernel} & \multicolumn{4}{c}{Strides of conv./upsampling layers} &  & \multicolumn{2}{c}{Encoded layer (L4)} &  & \multicolumn{2}{c}{Results} \\ \cmidrule{4-7} \cmidrule{9-10} \cmidrule{12-13} 
 &  &  & L1/8 & L2/7 & L3/6 & L4/5 &  & Shape ($\times$32) & Size &  & AIC & Acc. \\
\midrule
\multirow{3}{*}{CAE (Square)} & 1 & (5, 5) & (2, 2) & (2, 2) & (2, 2) & (2, 2) && (8, 8) & 2048 && \textbf{1820} & \textbf{73.3} \\
& 2 & (5, 5) & (2, 2) & (2, 2) & (2, 2) & (4, 4)  && (4, 4)  & 512 && 1925 & 66.7 \\
& 3 & (5, 5) & (2, 2) & (2, 2) & (4, 4) & (4, 4) && (2, 2)  & 128 && 1958 & 66.7 \\
\midrule
\multirow{4}{*}{CAE (Tall)} & 4 & (5, 3) & (2, 2) & (2, 2) & (2, 2) & (2, 4) && (8, 4)  & 1024 && \textbf{1609} & \textbf{73.3}  \\
& 5 & (5, 3) & (2, 2) & (2, 2) & (2, 4) & (2, 4) && (8, 2)  & 512 && 1647 & 70.0 \\
& 6 & (5, 3) & (2, 2) & (2, 4) & (2, 4) & (2, 4) && (8, 1)  & 256 && 2361 & 63.3 \\
& 7 & (5, 3) & (2, 2) & (2, 4) & (2, 4) & (4, 4) && (4, 1)  & 128 && 2523 & 56.7 \\
\midrule
\multirow{4}{*}{CAE (Wide)} & 8 & (3, 5) & (2, 2) & (2, 2) & (2, 2) & (4, 2) && (4, 8)  & 1024 && 1921 & 66.7 \\
& 9 & (3, 5) & (2, 2) & (2, 2) & (4, 2) & (4, 2) && (2, 8)  & 512 && 1866 & 73.3 \\
& 10 & (3, 5) & (2, 2) & (4, 2) & (4, 2) & (4, 2) && (1, 8)  & 256 && 1395 & 83.3 \\
& 11 & (3, 5) & (2, 2)  & (4, 2) & (4, 2) & (4, 4) && (1, 4)  & 128 && \textbf{1298} & \textbf{83.3} \\
\midrule
PK08 & 12 & -- & -- & -- & -- & -- & -- & -- & -- && \textbf{2388} & \textbf{53.3} \\
TEMP & 13 & -- & -- & -- & -- & -- & -- & -- & -- && 2692 & 40.0 \\
MFCC & 14 & -- & -- & -- & -- & -- & -- & -- & -- && 2703 & 46.7 \\
\bottomrule
\end{tabular}
\caption{Details of the CAEs and results for 14 feature sets. CAEs differ in the kernel shape of L1 and L8, and the shape of the encoded layer (determined by strides). Results are given in terms of \emph{i)} the LMER model fit (AIC), and \emph{ii)} the percentage of imitated drum sounds for which the $rating \sim\ distance$ slope is significantly less than 0 ($\alpha < 0.05$). Note: lower AIC = better model fit.}
\label{tab:results}
\end{table*}

\subsection{Linear mixed effect regression modelling}

For a given feature set, distance is measured between each of the 420 imitations and their respective 6 within-class sounds, giving 2520 distance values. We use Euclidean distance in keeping with the PK08 baseline method, and the distances for each feature set are normalised between 0--1. Linear mixed effect regression (LMER) models are then fitted for predicting the ratings from the distances. LMER is well suited to this task given that all listeners did not provide ratings for all imitations but only a randomly-selected set of 28 imitations (giving an unbalanced dataset). In addition, it allows us to include the dependencies between ratings for each listener and imitated sound. 

Maximum likelihood parameters for the models are estimated using the lme4 package in R \cite{bates2014}. The general model is fitted with rating $y_{ijk}$ as the dependent variable for each rating $i$, random intercepts for each listener $k$, and fixed effects of distance $x_{ij}$ and imitated sound $j$, with an interaction term between distance and imitated sound. The model is given by:

\begin{equation}
y_{ijk} = \nu_j + \beta_{1j} x_{ij} + \gamma_k + \epsilon_{ijk}  
\label{eq:1}
\end{equation}

\noindent where $\beta_{1j}$ is the slope of rating over distance for a given instance of $j$, and $\gamma_k$ is the random intercept for a given listener $k$. We note that model analysis showed heteroskedasticity in the residuals. Parameter estimates were therefore compared to those from robust models \cite{koller2016robustlmm}, and no major differences were found. As such the non-robust models were used for the analysis.  

Wald 95\% confidence intervals (CIs) were then calculated for the slope of each interaction ($\beta_{1j}$). For imitated sounds where the upper CI for $\beta_{1j} < 0$, we can infer the slope is significantly below 0 ($\alpha<0.05$). This indicates that the feature set is a good predictor for the imitated sound in question.  

The performance of each feature set is evaluated using two metrics: The percentage of imitated sounds for which $\beta_{1j}$ is significantly below 0 (accuracy); and Akaike's information criterion (AIC), which gives a measure of model fit (note: lower AIC = better model fit). An ideal feature set would have a significantly negative $\beta_{1j}$ (perfect predictor = -1.0) for all 30 imitated sounds, and be a good fit to the rating data given the model in Eq.\ \ref{eq:1}. 

\section{Results and Discussion}

\begin{figure}
	\begin{center}
		\includegraphics[width=\columnwidth]{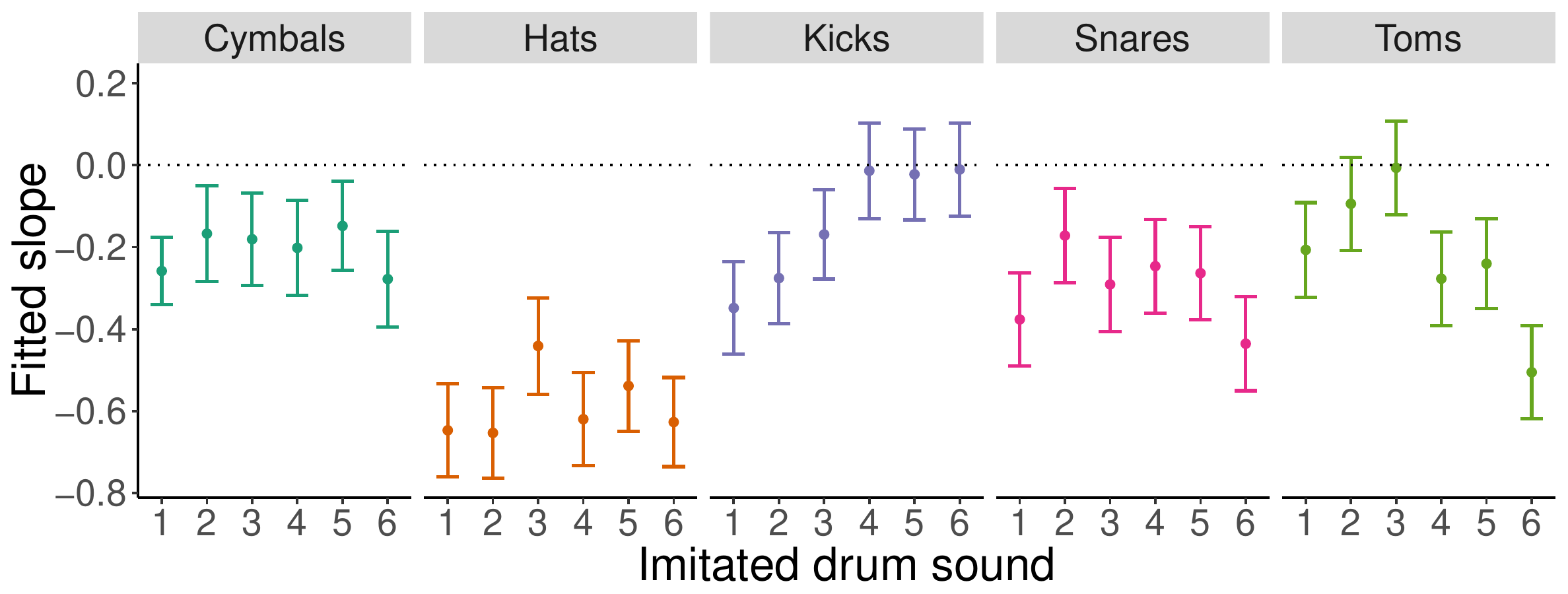}
		\caption{Slope estimates (with 95\% CIs) for the LMER model fitted on the performing feature set (11). A negative slope indicates a decrease in perceptual similarity with an increase in distance, i.e.\ sounds for which the feature set performs well.}
		\label{fig:slopes}
	\end{center}
\vspace{-6mm}
\end{figure}

The results are given in Table \ref{tab:results}. The encoded features from all CAEs outperform the baseline feature sets. The LMER model from the best performing feature set (11) gives fitted slopes for $rating \sim\ distance$ that are significantly less than 0 ($\alpha < 0.05$) for 83.3\% (25/30) of the imitated sounds, and has the lowest AIC. This shows the feature set is generally a good predictor of perceptual similarity between the vocal imitations and imitated sounds tested here, and has the best fitting LMER model. 

Interestingly, preservation of the temporal resolution is more important than spectral resolution for our task: for CAEs wide in time and narrow in frequency (8--11) performance improves as the size of the encoded layer decreases. This indicates redundancy in the spectral information: encoded shapes with spectral dimensions $>1$ have an adverse effect on performance. The similarity ratings are only for sounds in the same class (e.g.\ kick, snare etc.), and we expect high spectral similarity within each class. As such, overall energy differences in time may be more salient than the spectral distribution, providing the cues used by listeners when giving the ratings. This hypothesis is supported by comparing the square and tall CAEs: where reducing the size of the time dimension decreases performance. However there is also some redundancy in the temporal information, as can be seen comparing feature sets 10 and 11. As a post-hoc analysis we tested variants of CAE 11 using smaller encoded kernel shapes: ($1,2$) and ($1,1$), and found a decrease in performance below ($1,4$). This effect can also be seen in models 4--7, where performance decreases as width is reduced from 4 to 1. 

Regarding the baseline features, both TEMP and MFCC show similarly poor performance in terms of AIC (MFCC performs slightly better in terms of accuracy). This indicates that although the \emph{learned} temporal features appear to be most important for our task, the 5 heuristic temporal features are not sufficient to capture the salient cues used by listeners. The benefits of learned features over MFCCs concur with previous work \cite{zhang2015retrieving}, however we see greater disparity in performance. This may be specific to the sounds used in the evaluation (in \cite{zhang2015retrieving} a much wider range of sounds was used). The improved performance of PK08 compared to the other baselines indicates that this measure is somewhat transferable to vocalised drum sounds, although still only achieving an accuracy of 53\%.

Further analysis of the LMER model for the best performing feature set (11) shows the individual slopes for each drum sound (Fig. \ref{fig:slopes}). Here we observe considerable variation between the imitated sounds. In particular, we note that the 5 sounds for which the upper CI crosses 0 (3 kicks and 2 toms) are all pitched (although they are not the only pitched sounds in the dataset, indeed, all the toms are pitched). This suggests that reducing the size of the encoded spectral shape to 1 may work best over all the drum sounds used here, however the predictions for some pitched sounds suffer as a result.    

Finally, we note the slopes, although generally below 0, do not approach -1. Listener rating data is inherently noisy, and the concordance amongst listeners varies across the sounds. As such, there will clearly be a glass ceiling for performance, and a perfect model fit would not be useful for a real world application of the LMER model. Indeed, a perfect model fit is not desirable if one is interested in generalisability of the fitted LMER model.

\vspace{-1mm}

\section{Conclusions and Future Work}

In this paper we apply convolutional auto-encoders (CAEs) to query by vocalisation (QBV) for drum sound retrieval. We present a novel evaluation using perceptual similarity ratings between vocal imitations and the imitated drum sounds, providing insight into how learned features perform at predicting these ratings. Specifically, we compare CAEs that differ in both the size and shape of the encoded layer, in terms of the spectral and temporal dimensions. \\
Our experiments show that CAEs outperform 3 sets of heuristic features by a considerable margin. Furthermore, we show that reducing the size of the encoded layer height (frequency) increases the predictive power of the learned features, yet reducing the width (time) has the opposite effect. This finding is partly unexpected given that drum sounds generally have a similar overall temporal envelope (attack followed by a decay), however understandable given that we compare within-class sounds (e.g.~kick, snare etc.), which are also likely to share similar spectral distributions. For future work we would like to investigate more fine-grained morphological features to represent the temporal evolution that appears to be so important here. In addition we would like to investigate the generalisability of the best performing fitted LMER model to other QBV tasks, to determine how a model fitted on one set of sounds and similarity ratings performs given a larger sound library, as might be used in a typical music production environment.  

\vspace{-2mm}

\section{Acknowledgements}
This work is supported by EPSRC grants for the Media and Arts Technology Doctoral Training Centre (EP/G03723X/1) and FAST IMPACt project (EP/L019981/1). 

\bibliographystyle{IEEEbib}
\bibliography{ref}

\end{document}